\documentclass[acmsmall]{acmart}

\AtBeginDocument{%
  \providecommand\BibTeX{{%
    \normalfont B\kern-0.5em{\scshape i\kern-0.25em b}\kern-0.8em\TeX}}}

\usepackage{amsmath}
\renewcommand{\vec}[1]{\boldsymbol{#1}}
\usepackage{amsfonts}
\usepackage{subfigure}
\usepackage{multirow}
\usepackage{bm}
\usepackage{graphicx}
\usepackage{booktabs}
\usepackage{wrapfig}

\setcopyright{acmlicensed}
\acmJournal{TOIS}
\acmYear{2020}
\acmVolume{38}
\acmNumber{3}
\acmArticle{22}
\acmMonth{5}
\acmPrice{15.00}
\acmDOI{10.1145/3382764}

\begin{document}

\title{Exploiting Cross-Session Information for Session-based Recommendation with Graph Neural Networks}

\author{Ruihong Qiu}
\affiliation{%
  \institution{The University of Queensland}
  \city{Brisbane}
  \country{Australia}
}
\email{r.qiu@uq.edu.au}

\author{Zi Huang}
\affiliation{%
  \institution{The University of Queensland}
  \city{Brisbane}
  \country{Australia}
}
\email{huang@itee.uq.edu.au}

\author{Jingjing Li}
\affiliation{%
  \institution{University of Electronic Science and Technology of China}
  \city{Chengdu}
  \country{China}
}
\email{lijin117@yeah.net}

\author{Hongzhi Yin}
\authornote{Corresponding author. Contributing equally with the first author.}
\affiliation{%
  \institution{The University of Queensland}
  \city{Brisbane}
  \country{Australia}
}
\email{h.yin1@uq.edu.au}

\thanks{The work has been supported by ARC Discovery Project (Grant No. DP190101985 and DP170103954) and National Natural Science Foundation of China (Grant No. 61628206 and 61806039)}

\renewcommand{\shortauthors}{Ruihong Qiu, et al.}

\begin{abstract}
Different from the traditional recommender system, the session-based recommender system introduces the concept of the \textit{session}, i.e., a sequence of interactions between a user and multiple items within a period, to preserve the user's recent interest. The existing work on the session-based recommender system mainly relies on mining sequential patterns within individual sessions, which are not expressive enough to capture more complicated dependency relationships among items. In addition, it does not consider the cross-session information due to the anonymity of the session data, where the linkage between different sessions is prevented. In this paper, we solve these problems with the graph neural networks technique. First, each session is represented as a graph rather than a linear sequence structure, based on which a novel \textbf{F}ull \textbf{G}raph \textbf{N}eural \textbf{N}etwork (FGNN) is proposed to learn complicated item dependency. To exploit and incorporate cross-session information in the individual session's representation learning, we further construct a \textbf{B}roadly \textbf{C}onnected \textbf{S}ession (BCS) graph to link different sessions and a novel Mask-Readout function to improve session embedding based on the BCS graph. Extensive experiments have been conducted on two e-commerce benchmark datasets, i.e., \textit{Yoochoose} and \textit{Diginetica}, and the experimental results demonstrate the superiority of our proposal through comparisons with state-of-the-art session-based recommender models.
\end{abstract}

\ccsdesc[500]{Information systems~Recommender systems}

\keywords{recommender system, session-based recommendation, graph neural networks}

\maketitle

\section{Introduction}
\label{intro}
The recommender system (RS) has achieved great success in various online commercial applications such as e-commerce and social media platforms. There are two widely studied branches of the RS, the content-based RS~\cite{Pazzani2007ContentBasedRS} and the collaborative filtering RS~\cite{Schafer2007,he2017neural}, both of which focus on learning the user's preference towards items from the user's historical interactions with items. Conventionally, these methods mainly aim to explore all the historical data rather than focus on the user's recent interactions with items. As a result, the shift of a user's preference along the time is always neglected.

Recently, the session-based recommender system (SBRS) attracts more and more attention from both academia and industry. A session in SBRS refers to a sequence of interactions between a user and multiple items within a period. Compared with the traditional RS, SBRS focuses on learning the user's latest preference to recommend the next item based on the user's current ongoing session. How to predict a user's latest preference to recommend the next item is the core task of the SBRS.

\begin{figure}[t]
    \centering
    \includegraphics[width=0.4\linewidth]{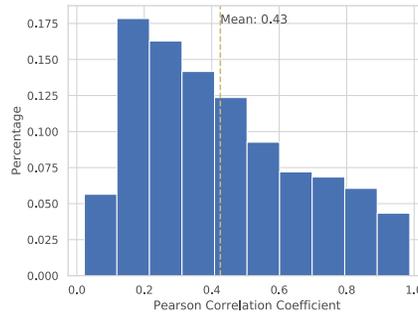}
    \vspace{-0.4cm}
    \caption{The distribution of Pearson correlation coefficient between sessions sharing at least one item on Diginetica dataset. The average Pearson correlation coefficient is $0.43$.}
\label{fig:digi5}
\end{figure}

Because of the nature of SBRS, most existing approaches on SBRS, such as GRU4REC~\cite{hidasi2015session}, NARM~\cite{li2017neural} and ATEM~\cite{WangHCHL018}, only consider the current session and treat it as a short sequence. They mainly adopt a recurrent neural network (RNN) or its variants to learn sequential patterns of items from individual sessions and the representation of a user's short-term preferences. In the procedure of generating the session sequence representation, most recent work~\cite{Liu18STAMP,wu2018session} makes use of the attention mechanism to differentiate the long-term and short-term preference. However, these approaches encounter two limitations. First, the linear sequential structure adopted by these methods is not expressive enough to represent and model the complicated non-sequential dependency relationships of items. Second, these approaches do not take cross-session information into account. An individual session tends to be very short, and the data of interactions from other sessions has a great potential to alleviate the data sparsity issue and to improve the recommendation accuracy.
As shown in Fig.~\ref{fig:digi5}, the distribution of Pearson correlation coefficient between sessions sharing at least one item indicates there is a strong positive correlation of sessions on benchmark datasets, e.g., Diginetica\footnote{http://cikm2016.cs.iupui.edu/cikm-cup/}. In such situation, it is inspired that the cross-session information can be helpful to tackle the data sparsity issue. Although some more recent work~\cite{hrm15,dream16,anam18,Hu18neural,QuadranaKHC17} attempts to leverage the cross-session information to improve the performance of session-based recommendation, all of them assume that the user ID of each session is available so that they can link the sessions that belong to the same user. For example, BINN~\cite{LiZLHMC18} studies long-term user preferences by applying bi-directional LSTMs (Bi-LSTMs) to the user's whole historical sessions. Unfortunately, they are inapplicable to the anonymous session data without user ID information.

Due to the above limitations of existing session-based recommendation methods, we aim to address the following two problems in this work: (1) how to model the complicated item dependency relationships to improve the item embedding and session embedding; (2) how to exploit and incorporate the cross-session information into the current session in the setting of anonymous sessions.

To address the first problem, we proposed the model \textbf{F}ull \textbf{G}raph \textbf{N}eural \textbf{N}etwork (FGNN) in the previous work~\cite{qiu2019rethinking} to learn complicated dependency relationships in individual sessions. Specifically, we first represent each session by a graph, which is more expressive than a sequence structure, and then apply a graph neural network to a session graph to learn both item and session embeddings. FGNN consists of two modules: (1) a weighted graph attention layer (WGAT) is proposed to encode the information among the nodes in the session graph into the item embeddings; (2) after obtaining the item embeddings, a Readout function, which learns to determine an appropriate item dependency, is designed to aggregate these embeddings to generate a graph level representation, i.e., the session embedding. At the final stage, the FGNN outputs the ranked recommendation list according to the score by comparing the session embedding with the embeddings of all items in the item set.

Although we have studied how to represent and model the complicated dependency relationships of items within individual sessions in our previous work~\cite{qiu2019rethinking}, the cross-session information was not considered. In this paper, we propose a new approach to expand the current session graph with other sessions, and the expanded session graph is named as the Broad Connected Session (BCS) graph, where the cross-session information is incorporated. To generate the effective embedding for each BCS graph, we propose a novel Mask-Readout function to aggregate the item embeddings with more attention on items from the original session graph, avoiding the learned session embedding being distracted by the cross-session information.

To sum up, this paper focuses on the session-based recommendation task. In the previous work~\cite{qiu2019rethinking}, we have demonstrated our preliminary study of learning the complicated dependency relationships of items. This paper extends the previous work by investigating the cross-session information with in-depth performance analysis. In detail, this paper makes the following new contributions:
\begin{itemize}
    \item A BCS graph is introduced for converting a session into a graph with the cross-session information.
    \item Based on the FGNN model, we propose Mask-Readout to generate the session embedding in the situation when cross-session information is incorporated.
    \item More extensive experiments are conducted to evaluate the performance of the BCS graph and the Mask-Readout.
    \item Comprehensive in-depth analysis is demonstrated along with a detailed review of related work.
\end{itemize}

The remainder of the paper is organized as follows: In Section~\ref{related-work}, the related work is reviewed in detail, followed by the description of the preliminaries of GNN in Section~\ref{preliminaries}. In Section~\ref{method}, the detail of the proposed model is presented. At last, extensive experiments are conducted and analyzed to verify the efficacy of the proposed model in Section~\ref{exp-init}.

\section{Related Work}
\label{related-work}
In this section, we firstly review some related work about the general recommender system (RS) in Section~\ref{rs} and the session-based recommender system (SBRS) in Section~\ref{sbrs}. At last, we will describe graph neural networks (GNN) for the node representation learning and graph classification problems in Section~\ref{gnn}.

\subsection{General Recommender System}
\label{rs}
The most popular method in recent years for the general recommender system is collaborative filtering (CF), which represents the user interest based on the whole history. For example, the famous shallow method, Matrix Factorization (MF)~\cite{koren2009matrix} factorizes the whole user-item interaction matrix with latent representation for every user and item. With the prevalence of deep learning, neural networks are widely used. Neural collaborative filtering (NCF)~\cite{he2017neural,chen-2019-joint} proposes to use the multilayer perceptron to approximate the matrix factorization process. More subsequent work extends the incorporation of different deep learning tools, for instance, convolutional neural networks~\cite{he2018outer}, knowledge graph~\cite{WangZWZLXG19} and zero-shot learning and domain adaptation~\cite{Li19from,li2019both}. To make use of the text information, Guan et al.~\cite{GuanCHZZPC19} proposed to use the attention to learn the relative importance of the reviews. These methods all depend on the identification of users and the whole record of interactions for every user. However, for many modern commercial online systems, the user information is anonymous, which leads to the failure of these CF-based algorithms.

\subsection{Session-based Recommender System}
\label{sbrs}
The research on the session-based recommender system (SBRS) is a sub-field of RS. Compared with RS, SBRS takes the user's recent user-item interactions into consideration rather than requiring all historical actions. SBRS is based on the assumption that the recent choice of items can be viewed as the recent preference of a user.

{\bf Sequential recommendation} is based on the Markov chain model~\cite{shani2005mdp,zimdars2001using,WeiqingTPM,ChenSeq}, which learns the dependency of items of a sequence data to predict the next click. Using probabilistic decision-tree models, Zimdars et al.~\cite{zimdars2001using} proposed to encode the state of the dependency relationships of the item sequential dependency. Shani et al.~\cite{shani2005mdp} made use of a Markov
Decision Process (MDP) to compute the probability of recommendation with the dependency probability between items.

{\bf Deep learning models} are popular recently with the boom of recurrent neural networks~\cite{hochreiter1997long,chung2014empirical,Chen19AIR,Wang16SPORE,Sun2019what,Sun20where}, which is naturally designed for processing sequential data. Hidasi et al.~\cite{hidasi2015session} proposed the GRU4REC, which applies a multi-layer GRU~\cite{chung2014empirical} to simply treat the data as time series. Based on the RNN model, some work makes improvements on the architectural choice and the objective function design~\cite{hidasi2018recurrent,tan2016improved}. In addition to RNN, Jannach and Ludewig~\cite{jannach2017recurrent} proposed to use the neighborhood-based method to capture co-occurrence signals. Incorporating content features of items, Tuan and Phuong~\cite{tuan20173d} utilized 3D convolutional neural networks to learn more accurate representations. Wu et al.~\cite{wu2017session} propose a list-wise deep neural network model to train a ranking model. Some recent work uses the attention mechanism to avoid the time order. NARM~\cite{li2017neural} stacks GRU as the encoder to extract information and then a self-attention layer to assign a weight to each hidden state to sum up as the session embedding. To further alleviate the bias introduced by time series, STAMP~\cite{Liu18STAMP} entirely replaces the recurrent encoder with an attention layer. SR-GNN~\cite{wu2018session} applies a gated graph network~\cite{li2015gated} as the item feature encoder and a self-attention layer to aggregate item features together as the session feature. FWSBR~\cite{Hwangbo_2019} is proposed to integrate the fashion feature to make sustainable session-based recommendation. SSRM~\cite{guo2019streaming} considers a specific user's historical sessions and applies the attention mechanism to combine them. Although the attention mechanism can proactively ignore the bias introduced by the time order of interactions, it considers the session as a totally random set.

{\bf Cross-session information} is considered in SBRS when the user ID is available~\cite{hrm15,dream16,anam18,Hu18neural,QuadranaKHC17}. All these methods require the identified user to build connections between sessions. Quadrana et al.~\cite{QuadranaKHC17} proposed HRNN to apply a recurrent architecture to aggregate the average feature of all sessions from the user's history. Bai et al.~\cite{anam18} proposed ANAM, which uses an attention model to combine different sessions. The user ID is the only linkage between different sessions in these methods. However, there is no record of user ID in many online systems, which makes it impossible to access to other sessions in this way.

\subsection{Graph Neural Networks}
\label{gnn}
In recent years, GNN attracts much interest in the deep learning community. Inspired by the embedding learning method Word2Vec~\cite{mikolov2013distributed}, DeepWalk~\cite{perozzi2014deepwalk} learns node embeddings by randomly sampling neighboring nodes and predicting the joint probabilities of these neighbors. With different learning objective functions and sampling strategies, LINE~\cite{tang2015line} and Node2Vec~\cite{grover2016node2vec} are the most representative algorithms of the unsupervised learning on the graph. Initially, GNN is applied to the simple situation on directed graphs~\cite{gori2005new,scarselli2009graph}. In recent years, many GNN methods~\cite{kipf2017semi,velickovic2018graph,hamilton2017inductive,li2015gated,xu2018how} work under the mechanism that is similar to message passing network~\cite{gilmer2017neural} to compute the information flow between nodes via edges. Additionally, the graph level feature representation learning is essential for graph level tasks, for example, graph classification and graph isomorphism~\cite{xu2018how,li2019graph}. Set2Set~\cite{vinyals2015order} assigns each node in the graph a descriptor as the order feature and uses this re-defined order to process all nodes. SortPool~\cite{zhang2018end} sorts the nodes based on their learned feature and uses a normal neural network layer to process the sorted nodes. DiffPool~\cite{ying2018hierarchical} designs two sets of GNN for every layer to learn a new dense adjacent matrix for a smaller size of the densely connected graph.

\section{Preliminaries}
\label{preliminaries}
In this section, we introduce how GNN works on the graph data. Let $G(V,E)$ denote a graph, where $v\in V$ is the node set with node feature vectors $\vec X_v$ and $e\in E$ is the edge set. There are two commonly popular tasks, e.g., \textit{node classification} and \textit{graph classification}. In this work, we focus on graph classification because our purpose is to learn a final embedding for the session rather than single items. For the \textit{graph classification}, given a set of graphs $\{G_1,\ldots,G_N\}\subseteq \mathcal{G}$ and the corresponding labels $\{y_1,\ldots,y_N\}\subseteq \mathcal{Y}$, we aim to learn a representation of the graph $\vec h_G$ to predict the graph label, $y_G=g(\vec h_G)$.

GNN makes use of the structure of the graph and the feature vectors of nodes to learn the representation of nodes or graphs. In recent years, most GNN work by aggregating information from neighboring nodes iteratively. After $k$ iterations of the update, the final representations of the nodes capture the structural information as well as the node information within $k$-hop neighbor. The procedure can be formed as
\begin{equation}
    \vec a_v^{(k)}=\text{Agg}(\{\vec h_u^{(k-1)},u\in N(v)\}), \vec h_v^{(k)}=\text{Map}(\vec h_v^{(k-1)},\vec a_v^{(k)}),
\end{equation}
where $\vec h_v^{(k)}$ is the feature vector for node $v$ in the $k$th layer. For the input $\vec h_v^{0}$ to the first layer, the feature vectors $\vec X_v$ are passed in. Agg and Map are two functions that can be defined in different forms. Agg serves as the aggregator to aggregate features of neighboring nodes. A typical characteristic of Agg is permutation invariant. Map is a mapping to transform the self information and the neighboring information to a new feature vector.

For the graph classification, a Readout function aggregates all node features from the final layer of the graph to generate a graph level representation $\vec h_G$:
\begin{equation}
    \vec h_G=\text{Readout}(\{\vec h_v^{(k)},v\in V\}),
\end{equation}
where the Readout function needs to be permutation invariant as well.

\section{Method}
\label{method}
\begin{figure}[t]
    \centering
    \includegraphics[width=\linewidth]{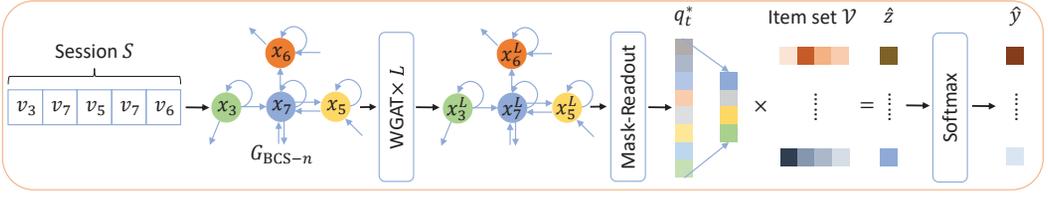}
    \vspace{-0.4cm}
    \caption{The pipeline of FGNN. The input to the model is a BCS graph $G_{\text{BCS}-n}$, which is converted from the input session $S$. $\vec x$ is the feature vector of nodes in $G_{\text{BCS}-n}$. $L$ layers of WGAT serve as the encoder of node features for $G_{\text{BCS}-n}$. After being processed by WGAT, the session graph now contains different semantic node representations $x^L$ but with the same structure as the input session graph. The Mask-Readout function is applied to generate a session embedding based on the learned node features. Compared with other items in the item set $\mathcal{V}$, a recommendation score $\hat{\vec y_i}$ is finally generated.}
    \label{fig:whole}
\end{figure}

In this section, we describe our \textbf{F}ull \textbf{G}raph \textbf{N}eural \textbf{N}etwork (FGNN) model in detail. The overview of the FGNN model is shown in Fig.\ \ref{fig:whole}.

\subsection{Problem Definition and Notation}
\label{problem}
The purpose of an SBRS is to predict the next item that matches the current anonymous user's preference based on the interactions within the session. In the following, we give out the definition of the SBRS problem.

In an SBRS, there is an item set $\mathcal{V}=\{v_1,v_2,v_3,\ldots,v_m\}$, where all items are unique and $m$ denotes the number of items. A session sequence from an anonymous user is defined as an order list $\mathcal{S}=[v_{s,1},v_{s,2},v_{s,3},\ldots,v_{s,n}]$, $v_{s,*} \in \mathcal{V}$. $n$ is the length of the session $\mathcal{S}$, which may contain duplicated items, $v_{s,p}=v_{s,q}$, $p, q < n$. The goal of our model is to take an anonymous session $\mathcal{S}$ and the cross-session information as input, and predict the next item $v_{s,n+1}$ that matches the current anonymous user's preference.

\subsection{BCS Graph}
\subsubsection{Basic Session Graph}
\label{session-graph}
As shown in Fig.\ \ref{fig:whole}, at the first stage, the session sequence is converted into a session graph for the purpose to process each session via GNN. Because of the natural order of the session sequence, we transform it into a weighted directed graph, $G_s=(V_s,E_s)$, $G_s \in \mathcal{G}$, where $\mathcal{G}$ is the set of all session graphs. In the session graph $G_s$, the node set $V_s$ represents all nodes, which are items $v_{s,n}$ from $S$. For every node $\vec v$, the input feature is the initial embedding vector $\vec x$. The edge set $E_s$ represents all directed edges $(v_{s,n-1},v_{s,n},w_{s,(n-1)n})$, where $v_{s,n}$ is the click of item after $v_{s,n-1}$ in $S$, and $w_{s,(n-1)n}$ is the weight of the edge. The weight of the edge is defined as the frequency of the occurrence of the edge within the session. For convenience, in the following, we use the nodes in the session graph to stand for the items in the session sequence. For the self-attention used in WGAT introduced in Section~\ref{wgat}, if a node does not contain a self loop, it will be added with a self loop with a weight 1. Based on our observation of our daily life and the datasets, it is common for a user to click two consecutive items a few times within the session. After converting the session into a graph, the final embedding of $S$ is based on the calculation on this session graph $G_s$.

\subsubsection{BCS Graph}
\label{bcs-graph}
In this extension, the cross-session information extraction is another target. Instead of simply using the individual session to build a basic session graph as we did in the previous version, we want to enable the session graph to represent more information from the dataset. Therefore, a Broadly Connected Session graph is built upon the base of a session graph introduced in Section~\ref{session-graph} with more nodes and edges extracted from other sessions. The purpose of augmenting the session graph to a BCS graph is to incorporate cross-session information into the individual session representation learning procedure.

In the original session graph, nodes and edges are items and dependency respectively occurring in the individual session. If we want to make use of the cross-session information, we need to define the proper connection between different sessions. This is simply because only relevant sessions are meaningful for the recommendation task. In the traditional session sequence setting, it is difficult to define the relevance between sessions. Instead, for our session graph setting, it is straightforward to define the relevance without computing any similarity of different sessions.

\begin{figure}[t]
    \centering
    \includegraphics[width=\linewidth]{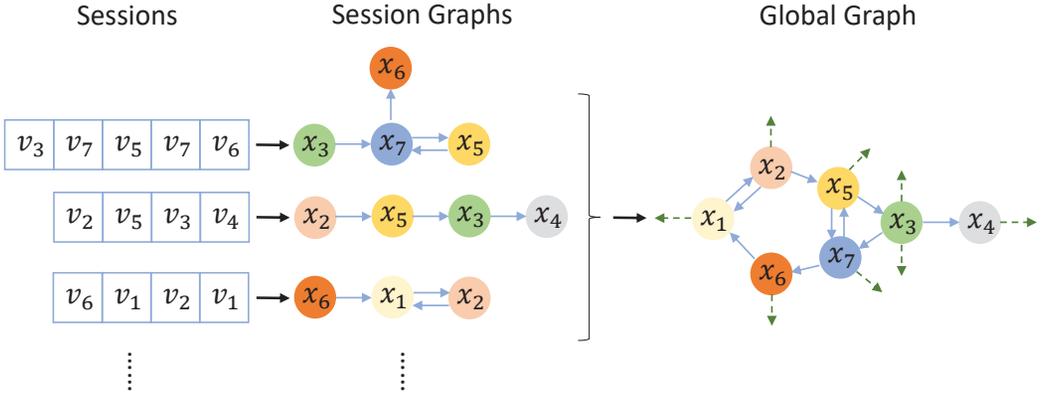}
    \vspace{-0.4cm}
    \caption{Generation of the global graph. Every single session is firstly converted into basic session graphs. Based on the re-occurrence of items, different session graphs can be united as the global graph.}
    \label{fig:full}
\end{figure}

First of all, before building a single session graph for every session, we gather all sessions and build a large graph to unify all of them. We refer to this large graph, which contains all sessions, as \textit{global graph}, $G_\text{full}=(V,E)$, where $V$ contains all items appearing in the training sessions and $E$ contains all dependency. The detailed generation procedure of the $G_\text{full}$ is presented in Fig.\ \ref{fig:full}. The weights of edges are defined as the way introduced in the basic session graph.

\begin{figure}[t]
    \centering
    \subfigure[A part of the global graph.]{
    \label{fig:sample-full}
    \includegraphics[width=0.4\linewidth]{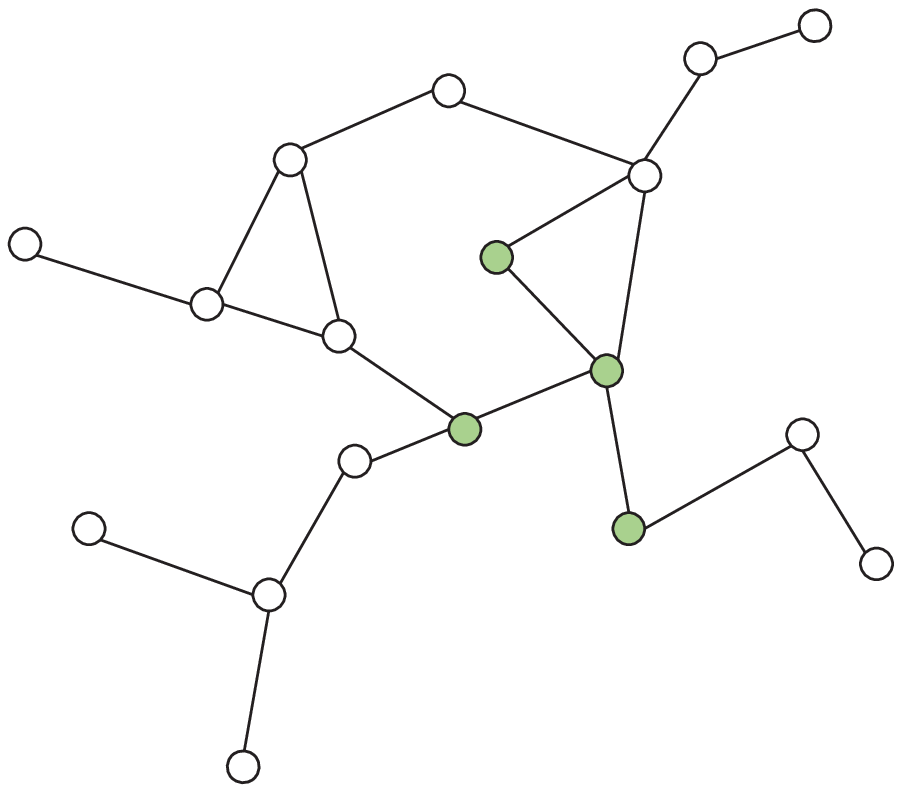}
    }
    \subfigure[The hierarchical structure of sampled $G_{\text{BCS}-n}$.]{
    \label{fig:sample-bcs}
    \includegraphics[width=0.4\linewidth]{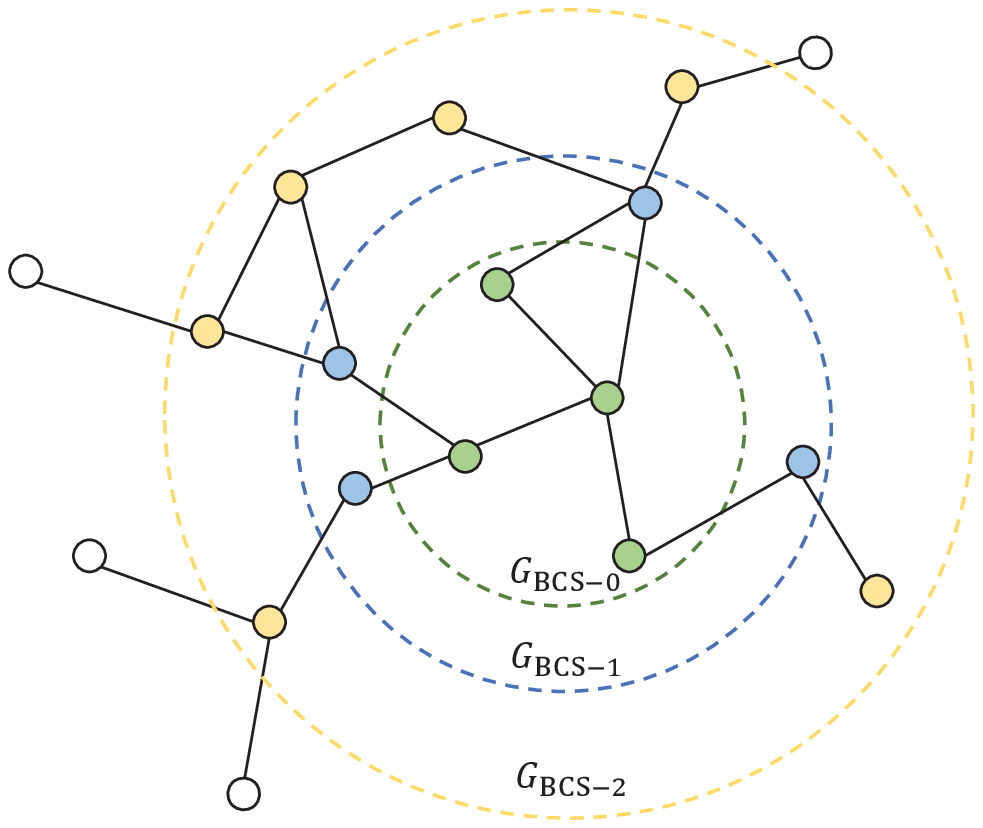}
    }
    \vspace{-0.4cm}
    \caption{An example of how to sample a BCS graph with $n$-hop neighbors from a global graph generated in the way described in Fig.\ \ref{fig:full}.We omit the directed graph setting here to make the demonstration clearer. The actual sampling procedure for the directed situation is totally the same.~\subref{fig:sample-full} A part of the global graph is shown here. Green nodes are represented as the input session graph, which does not include any neighbors of the nodes appearing in the session.~\subref{fig:sample-bcs} The layer with green nodes are actually the $G_{\text{BCS}-0}$. Furthermore, if we want to sample the $G_{\text{BCS}-1}$ of the input session, the resulting graph is the layer circled with blue. $G_{\text{BCS}-1}$ consists of $G_{\text{BCS}-0}$ and the first order neighbors of nodes in the input session. Similarly, $G_{\text{BCS}-2}$, which is circled with yellow, consists of $G_{\text{BCS}-0}$ and the first and second order neighbors of nodes in the input session.}
    \vspace{-0.5cm}
\label{fig:sample}
\end{figure}

\begin{figure}[t]
    \centering
    \subfigure[A session graph.]{
    \label{fig:session-graph}
    \includegraphics[width=0.25\linewidth]{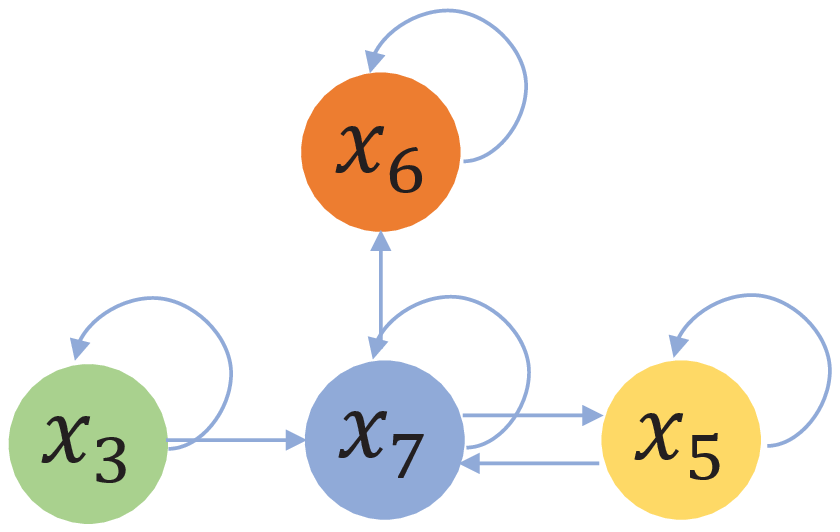}
    }
    \subfigure[The computation for the second-layer feature $\vec x''_6$.]{
    \label{fig:2layer}
    \includegraphics[width=0.25\linewidth]{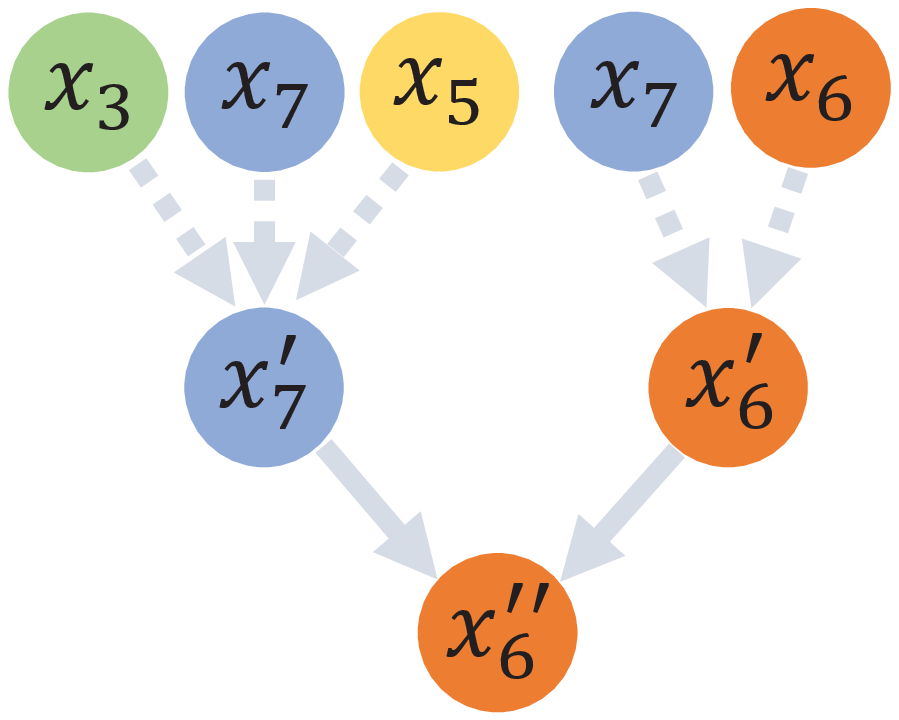}
    }
    \vspace{-0.4cm}
    \caption{An example of how to compute a node representation of a two-layer GNN. The original session sequence is the same as the input in Fig.\ \ref{fig:whole}.~\subref{fig:session-graph} The session graph is added with self-loop to every node. $\vec x_i$ is the input feature for the corresponding node $v_i$.~\subref{fig:2layer} The computation of the second-layer feature $\vec x''_6$ is based on all the first and second order neighboring nodes of $v_6$. The first order neighbors are $v_7$ and $v_6$ itself. The second order neighbors of $v_6$ are the first order neighbors of $v_7$ and $v_6$, i.e., $v_3$, $v_7$ and $v_6$ for $v_7$, and $v_7$ and $v_6$ for $v_6$.}
    \vspace{-0.5cm}
\label{fig:propagation}
\end{figure}

When processing the individual session, $\mathcal{S}=[v_{s,1},v_{s,2},v_{s,3},\ldots,v_{s,n}]$, we need to sample a sub-graph from the global graph according to the items in $\mathcal{S}$. Without any modification, we can simply sample a basic session graph, $G_s=(V_s,E_s)$, where $V_s=\{v_i \in \mathcal{S}\}$, corresponding to $\mathcal{S}$ from the global graph $G_\text{full}$. The sampled basic session graph here has the same structure as the one built with the method introduced in Section~\ref{session-graph}. But the same structure fails to provide cross-session information to the original session graph setting. It is worth noting that a typical GNN layer can be stacked to multiple layers, which compute the feature of nodes based on their $n$-hop neighbors, $N_{v_i}$. Therefore, we can sample a larger graph including $n$-hop neighbors of items in $\mathcal{S}$ from the global graph. We refer to this larger graph as the Broadly Connected Session graph, $G_{\text{BCS}-n}=(V_{\text{BCS}-n}, E_{\text{BCS}-n})$, where $V_{\text{BCS}-n}=\{v_i,N_{v_i}| v_i \in \mathcal{S}\}$ and $N_{v_i}$ is the $n-$hop neighbors of node $v_i$. When $n=0$, which means that the BCS graph $G_{\text{BCS}-n}$ does not include any neighbors of nodes appearing in the session. In this special case, $G_{\text{BCS}-n}=G_s$. The precise sampling process of a $G_{\text{BCS}-n}$ is shown in Fig.\ \ref{fig:sample}. In the Fig.\ \ref{fig:sample-full}, the global graph is exactly the one built with the way we describe above; in the Fig.\ \ref{fig:sample-bcs}, the hierarchical structure of sampling different $n$-hop BCS graph is represented in different colors.

A BCS graph sampled from the global graph is expected to contain much more cross-session information. However, for some popular items, they can appear in a great number of sessions. If we sample their whole neighboring nodes, it will drastically increase the size of the BCS graph. Therefore, in the neighbor level, a sampling procedure performs random selection over neighboring nodes to control the scale of the BCS graph. The sampling is based on the edge weight, which indicates the popularity of the following node.

Compared to the basic session graph, the $G_{\text{BCS}-n}$ contains more nodes and edges based on the extra relationship between items from other sessions. Such extra information extracted from $G_\text{complete}$ actually provides cross-session information. There is no need to calculate the similarity between sessions for our method to determine what kind of information is relative.

\subsection{Weighted Graph Attentional Layer}
\label{wgat}
After obtaining the session graph, a GNN is needed to learn embeddings for nodes in a graph, which is the $\text{WGAT}\times L$ part in Fig.\ \ref{fig:whole}. In recent years, some baseline methods on GNN, for example, GCN~\cite{kipf2017semi} and GAT~\cite{velickovic2018graph}, are demonstrated to be capable of extracting features of the graph. However, most of them are only well-suited for unweighted and undirected graphs. For the session graph, weighted and directed, these baseline methods cannot be directly applied without losing the information carried by the weighted directed edges. Therefore, a suitable graph convolutional layer is needed to effectively convey information between the nodes in the graph.

In this paper, we propose a weighted graph attentional layer (WGAT), which simultaneously incorporates the edge weight when performing the attention aggregation on neighboring nodes. We describe the forward propagation of WGAT in the following. The information propagation procedures are shown in Fig.\ \ref{fig:propagation}. Fig.\ \ref{fig:2layer} shows an example of how a two-layer GNN calculates the final representation of the node $v_6$.

The input to a WGAT is a set of node initial features, the item embeddings, $\vec x=\{\vec x_0,\vec x_1,\vec x_2,\ldots \vec x_{n-1}\}$, $\vec x_i \in \mathbb{R}^d$, where $n$ is the number of nodes in the graph, and $d$ is the dimension of the embedding $\vec x_i$. After applying the WGAT, a new set of node features, $\vec x'=\{\vec x'_0,\vec x'_1,\vec x'_2,\ldots \vec x'_{n-1}\}$, $\vec x'_i \in \mathbb{R}^{d'}$, will be given out as the output. Specifically, the input feature vectors $\vec x^0_i$ of the first WGAT layer are generated from an embedding layer, whose input is the one-hot encoding of items,
\begin{equation}
\label{eq:embed}
    \vec x^0_i=\text{Embed}(v_i),
\end{equation}
where Embed is the embedding layer.

To learn the node representation via the complicated item dependency relationships within the graph structure, a self-attention mechanism for every node $i$ is used to aggregate information from its neighboring nodes $\mathcal{N}(i)$, which is defined as the nodes with edges towards the node $i$ (may contain $i$ itself if there is a self-loop edge). Because the size of the session graph is not huge, we can take the entire neighborhood of a node into consideration without any sampling. At the first stage, a self-attention coefficient $e_{ij}$ to determine how importantly the node $j$ will influence the node $i$ is calculated based on $\vec x_i$, $\vec x_j$ and $w_{ij}$,
\begin{equation}
\label{e-ij}
    e_{ij}=\text{Att}(\vec W\vec x_i,\vec W\vec x_j,w_{ij}),
\end{equation}
where Att is a mapping $\text{Att}: \mathbb{R}^{d} \times \mathbb{R}^{d} \times \mathbb{R}^{1} \to \mathbb{R}^{1}$ and $\vec W$ is a shared parameter which performs linear mapping across all nodes. As a matter of fact, the attention of a node $i$ can extend to every node, which is a special case the same as how STAMP makes the attention of the last node of the sequence. Here we restrict the range of the attention within the first order neighbors of the node $i$ to make use of the inherent structure of the session graph $S$. To compare the importance of different nodes directly, a softmax function is applied to convert the coefficient into a probability form across the neighbors and itself,
\begin{equation}
    \alpha_{ij}=\text{softmax}_j(e_{ij})=\frac{\text{exp}(e_{ij})}{\sum_{k\in \mathcal{N}(i)}\text{exp}(e_{ik})}.
\end{equation}

The choice of $att$ can be diversified. In our experiments, we use an MLP layer with the parameter $\vec W_{att} \in \mathbb{R}^{2d+1}$, followed by a LeakyRelu non-linearity unit with negative input slope $\alpha=0.2$
\begin{equation}
\label{alpha-ij}
    \alpha_{ij}=\frac{\text{exp}(\text{LeakyRelu}(\vec W_{att}[\vec W\vec x_i||\vec W\vec x_j||w_{ij}]))}{\sum_{k\in\mathcal{N}(i)}\text{exp}(\text{LeakyRelu}(\vec W_{att}[\vec W\vec x_i||\vec W\vec x_k||w_{ik}]))},
\end{equation}
where $||$ means concatenation of two vectors.

For every node $i$ in $G_s$, in a WGAT layer, all attention coefficients of their neighbors can be computed as (\ref{alpha-ij}). To utilize these attention coefficients, a linear combination for the corresponding neighbors is applied to update the features of the nodes.
\begin{equation}
\label{1head}
    \vec x'_i=\sigma(\sum\limits_{j\in\mathcal{N}(i)}\alpha_{ij}\vec W\vec x_j),
\end{equation}
where $\sigma$ is a non-linearity unit and in our experiments, we use the ReLU~\cite{nair2010rectified}.

As suggested in previous work~\cite{velickovic2018graph,vaswani2017attention}, the multi-head attention can help to stabilize the training of the self-attention layers. Therefore, we apply the multi-head setting for our WGAT.
\begin{equation}
\label{multihead}
    \vec x'_i=\mathop{\Arrowvert}\limits_{k=1}^K\sigma(\sum\limits_{j\in\mathcal{N}(i)}\alpha^k_{ij}\vec W^k\vec x_j),
\end{equation}
where $K$ is the number of heads and for every head, there is a different set of parameters. $\Arrowvert$ in (\ref{multihead}) stands for the concatenation of all heads. As a result, after the calculation of (\ref{multihead}), $\vec x'_i \in \mathbb{R}^{K{d'}}$.

Specifically, if we stack multiple WGAT layers, the final nodes feature will be shaped as $\mathbb{R}^{K{d'}}$ as well. However, what we expect is $\mathbb{R}^{d'}$. Consequently, we calculate the mean over all the heads of the attention results.
\begin{equation}
\label{mean}
    \vec x'_i=\sigma(\frac{1}{K}\sum\limits^K_{k=1}\sum\limits_{j\in\mathcal{N}(i)}\alpha^k_{ij}\vec W^k\vec x_j).
\end{equation}

Once the forward propagation of multiple WGAT layers has finished, we obtain the final feature vectors of all nodes, which is the item level embeddings. These embeddings will serve as the input of the session embedding computation stage that we detail below.

\subsection{Mask-Readout Function}
\label{cent-read}
The purpose of the Mask-Readout function is to generate a representation of the updated BCS graph based on the node features after the forward computation of the GNN layers. The Mask-Readout needs to learn the description of the item dependency relationships to avoid the bias of the time order and the inaccuracy of the self-attention on the last input item. For the convenience, some algorithms use simple permutation invariant operations, for example, $Mean, Max$ or $Sum$ over all node features. Although it is clear that these methods are simple and not going to violate the constraints of the permutation invariance, they can not provide a sufficient model capacity for learning a representative session embedding for the BCS graph. In contrast, Set2Set~\cite{vinyals2015order} is a graph level feature extractor that learns a query vector indicating the order of reading from the memory for an undirected graph. We develop our basic Readout function by modifying this method to suit the setting of the BCS graphs. The computation procedures are as follows:
\begin{equation}
\label{set2set}
    \vec q_t=\text{GRU}(\vec q^*_{t-1}),
\end{equation}
\begin{equation}
    e_{i,t}=f(\vec x_i,\vec q_t),
\end{equation}
\begin{equation}
    a_{i,t}=\frac{\text{exp}(e_{i,t})}{\sum_j\text{exp}(e_{j,t})},
\end{equation}
\begin{equation}
    \vec r_t=\sum\limits_ia_{i,t}\vec x_i,
\end{equation}
\begin{equation}
\label{q*t}
    \vec q^*_t=\vec q_t\Arrowvert \vec r_t,
\end{equation}
where $i$ indexes node $i$ in the session graph $G_s$, $\vec q_t$, $\vec q_t \in \mathbb{R}^d$, is a query vector which can be seen as the order to read $\vec r_t\in \mathbb{R}^d$ from the memory and GRU is the gated recurrent unit, which at the first
step takes no inputs and at the following steps, takes the former output $\vec q^*_{t-1}\in \mathbb{R}^{2d}$. $f$ calculates the attention coefficient $e_{i,t}$ between the embedding of every node $\vec x_i$ and the query vector $\vec q_t$. $a_{i,t}$ is the probabilistic form of $e_{i,t}$ after applying a softmax function over $e_{i,t}$, which is then used to a linear combination on the node embeddings $\vec x_i$. The final output $\vec q^*_t$ of one forward computation of the Readout function is the concatenation of $\vec q_t$ and $\vec r_t$.

Based on all node embeddings for a session graph, we follow Equation~\ref{set2set}$\sim$\ref{q*t} to obtain a graph level embedding that contains a query vector $\vec q_t$ in addition to the semantic embedding vector $\vec r_t$. The query vector $\vec q_t$ controls what to read from the node embeddings, which actually provides an order to process all nodes if we recursively apply the Readout function.

Generally, there is no restriction for the Readout function to calculate over all nodes in the BCS graph. However, based on the purpose of the SBRS, the current session is delegated to reflect the user's most recent preference rather than sessions from previous time or other users. Therefore, it is necessary to preserve the original session information within the BCS graph when generating the graph embedding. To achieve this goal, the Mask-Readout masks out all nodes out of the range of $G_{\text{BCS}-0}$.

\begin{figure}
    \centering
    \includegraphics[width=0.4\linewidth]{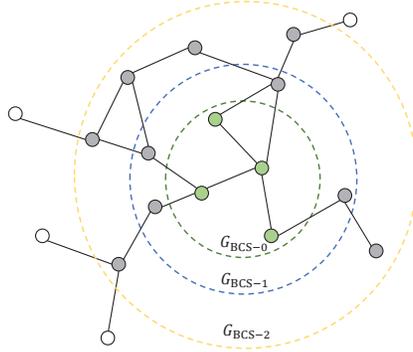}
    \vspace{-0.4cm}
    \caption{Mask-Readout. In the generation process of the graph embedding, Mask-Readout masks out all nodes (in gray) out of the range of $G_{\text{BCS}-0}$ (in green).}
    \vspace{-0.5cm}
    \label{fig:mask}
\end{figure}

As shown in Fig.~\ref{fig:mask}, the BCS session graph can vary in size with a different choice of the neighborhood. Under this situation, the basic Readout function calculates the session embedding based on all the nodes in the BCS graph, which fails to preserve the relative importance of the current session and other sessions. Therefore, the Mask-Readout function masks all the nodes out of the range of $G_{\text{BCS}-0}$ in the BCS graph to generate a precise session embedding.

\subsection{Recommendation}
\label{rec}
Once the graph level embedding $\vec q^*_t$ is obtained, we can use it to make a recommendation by computing a score vector $\hat{\vec z}$ for every item over the whole item set $\mathcal{V}$ with their initial embeddings in the matrix form,
\begin{equation}
\label{z}
    \hat{\vec z}=(\vec W_{out}\vec q^*_t)^T \vec X^0,
\end{equation}
where $\vec W_{out} \in \mathbb{R}^{d\times2d}$ is a parameter that performs a linear mapping on the graph embedding $\vec q^*_t$, the $T$ means the transformation on a matrix, and $\vec X^0$ is from the Equation~\ref{eq:embed}.

For every item in the item set $\mathcal{V}$, we can calculate a recommendation score and combine them together, we obtain a score vector $\hat{\vec z}$. Furthermore, we apply a softmax function over $\hat{\vec z}$ to transform it into the probability distribution form $\hat{\vec y}$,
\begin{equation}
\label{y}
    \hat{\vec y}=\text{softmax}(\hat{\vec z}).
\end{equation}

For the top-K recommendation, it is simple to choose the highest K probabilities over all items based on $\hat{\vec y}$.

\subsection{Objective Function}
\label{loss}
Since we already have the recommendation probability of a session, we can use the label item $v_{label}$ to train our model with the supervised learning method.

As mentioned above, we formulate the recommendation as a graph level classification problem. Consequently, we apply the multi-class cross entropy loss between $\hat{\vec y}$ and the one-hot encoding of $v_{label}$ as the objective function. For a batch of training sessions, we can have
\begin{equation}
    L=-\sum\limits_{i=1}^{l}\text{one-hot}(v_{label,i})\text{log}(\hat{\vec y_i}),
\end{equation}
where $l$ is the batch size we use in the optimizer.

In the end, we use the Back-Propagation Through Time (BPTT) algorithm to train the whole FGNN model.

\section{Experiments}
\label{exp-init}

In this section, we conduct experiments with the purpose to prove the efficacy of our proposed FGNN model by answering the following research questions:
\begin{itemize}
    \item \textbf{RQ1} Does the FGNN outperform other state-of-the-art SBRS methods? (in Section~\ref{rq1})
    \item \textbf{RQ2} How does the BCS graph help to make use of the cross-session information for anonymous sessions? (in Section~\ref{rq2})
    \item \textbf{RQ3} How does the WGAT work for the session-based recommendation problem? (in Section~\ref{rq3})
    \item \textbf{RQ4} How does the Readout function work differently from other graph level embedding methods? (in Section~\ref{rq4})
    \item \textbf{RQ5} How does the Mask-Readout function perform compared with the Readout function? (in Section~\ref{rq5})
\end{itemize}

In addition to the previous version, we introduce the BCS graph module to our FGNN model. As a result, we also emphasize on proving the effect of the BCS graph module in Section~\ref{rq2} in this version. Besides, there will be additional experiments involving the BCS graph in other sections.

In the following, we first describe the details of the basic setting of the experiments and afterwards, we answer the questions above by showing the results of the experiments.

\subsection{Datasets}
\label{datasets}
We follow the previous version to choose two representative real world e-commerce datasets i.e., \textit{Yoochoose}\footnote{https://2015.recsyschallenge.com/challenge.html} and \textit{Diginetica}, to evaluate our model.
\begin{itemize}
    \item \textit{Yoochoose} is used as a challenge dataset for RecSys Challenge 2015. It is obtained by recording click-streams from an e-commerce website within 6 months.
    \item \textit{Diginetica} is used as a challenge dataset for CIKM cup 2016. It contains the transaction data which is suitable for session-based recommendation.
\end{itemize}

\begin{table*}[t]
    \centering
    \caption{Statistic details of datasets.}
    \vspace{-0.3cm}
    \scalebox{0.8}{
    \begin{tabular}{lccccc}
         \toprule
         Dataset&Clicks&Train sessions&Test sessions&Items&Avg. length\\
         \midrule
         Yoochoose1/64&557248&369859&55898&16766&6.16\\
         Yoochoose1/4&8326407&5917746&55898&29618&5.71\\
         Diginetica&982961&719470&60858&43097&5.12\\
         \bottomrule
    \end{tabular}}
    \label{tab:datasets}
\end{table*}

For the fairness and the convenience of comparison, we follow~\cite{li2017neural,Liu18STAMP,wu2018session} to filter out sessions of length 1 and items which occur less than 5 times in each dataset respectively. After the preprocessing step, there are 7,981,580 sessions and 37,483 items remaining in \textit{Yoochoose} dataset, while 204,771 sessions and 43097 items in \textit{Diginetica} dataset. Similar to~\cite{wu2018session,tan2016improved}, we split a session of length $n$ into $n-1$ partial sessions of length ranging from $2$ to $n$ to augment the datasets. For the partial session of length $i$ in the session $S$, it is defined as $[v_{s,0},\ldots,v_{s,i-1}]$ with the last item $v_{s,i-1}$ as $v_{label}$. Following~\cite{li2017neural,Liu18STAMP,wu2018session}, for \textit{Yoochoose} dataset, the most recent portions $1/64$ and $1/4$ of the training sequence are used as two split datasets respectively.

\subsection{Baselines}
\label{baselines}
In order to prove the advantage of our proposed FGNN model, we compare FGNN with the following representative methods:
\begin{itemize}
    \item \textbf{POP} always recommends the most popular items in the whole training set, which serves as a strong baseline in some situations although it is simple.
    \item \textbf{S-POP} always recommends the most popular items for the individual session.
    \item \textbf{Item-KNN}~\cite{sarwar2001item} computes the similarity of items by the cosine distance of two item vectors in sessions. Regularization is also introduced to avoid the rare high similarities for unvisited items.
    \item\textbf{BPR-MF}~\cite{rendle2009bpr} proposes a BPR objective function which utilizes a pairwise ranking loss to train the ranking model. Following~\cite{li2017neural}, Matrix Factorization is modified to session-based recommendation by using mean latent vectors of items in a session.
    \item \textbf{FPMC}~\cite{rendle2010factorizing} is a hybrid model for the next-basket recommendation and it achieves state-of-the-art results. For anonymous session-based recommendation, following~\cite{li2017neural}, we omit the user feature directly because of the unavailability.
    \item \textbf{GRU4REC}~\cite{hidasi2015session} stacks multiple GRU layers to encode the session sequence into a final state. It also applies a ranking loss to train the model.
    \item \textbf{NARM}~\cite{li2017neural} extends to use an attention layer to combine all of the encoded states of RNN, which enables the model to explicitly emphasize on the more important parts of the input.
    \item \textbf{STAMP}~\cite{Liu18STAMP} uses attention layers to replace all RNN encoders in previous work to even make the model more powerful by fully relying on the self-attention of the last item in a sequence.
    \item \textbf{SR-GNN}~\cite{wu2018session} applies a gated graph convolutional layer~\cite{li2015gated} to obtain item embeddings, followed by a self-attention of the last item as \textbf{STAMP} does to compute the sequence level embeddings.
\end{itemize}

\subsection{Evaluation Metrics}
\label{eval}
For each time, a recommender system can give out a few recommended items and a user would choose the first few of them. To keep the same setting as previous baselines, we mainly choose to use top-20 items to evaluate a recommender system and specifically, two metrics, i.e., \textbf{R@20} and \textbf{MRR@20}. For more detailed comparison, top-5 and top-10 results are considered as well.

\begin{itemize}
    \item \textbf{R@K} (Recall calculated over top-K items). The R@K score is the metric that calculates the proportion of test cases which recommends the correct items in a top K position in a ranking list,
    \begin{equation}
        \text{R@K}=\frac{n_{hit}}{N},
    \end{equation}
    where $N$ represents the number of test sequences $S_{test}$ in the dataset and $n_{hit}$ counts the number that the desired items are in the top K position in the ranking list, which is named the $hit$. R@K is also known as the hit ratio.
    \item \textbf{MRR@K} (Mean Reciprocal Rank calculated over top-K items). The reciprocal is set to $0$ when the desired items are not in the top K position and the calculation is as follows,
    \begin{equation}
        \text{MRR@K}=\frac{1}{N}\sum\limits_{v_{label}\in S_{test}}\frac{1}{Rank(v_{label})}.
    \end{equation}
    The MRR is a normalized ranking of $hit$, the higher the score, the better the quality of the recommendation because it indicates a higher ranking position of the desired item.
\end{itemize}

\subsection{Experiments Setting}
\label{exp-set}
In the experiments, there are two types of building methods of the session graph. For the basic session graph, we directly make use of all nodes in the session to build the corresponding session graph. On the other hand, for the BCS graph, we sample the neighboring nodes according to the edge weight by restricting the number of a node's neighbors to 5 in our default setting. The neighboring sample rate will be discussed in Section~\ref{rq2}. We apply a three-layer WGAT and each with eight heads as our node representation encoder and three processing steps of our Readout function. The size of the feature vector of the item is set to 100 for every layer including the initial embedding layer. All parameters of the FGNN are initialized using a Gaussian distribution with a mean of 0 and a standard deviation of 0.1 except for the GRU unit in the Readout function, which is initialized using the orthogonal initialization~\cite{saxe2013exact} because of its performance on RNN-like units. We use the Adam optimizer with the initial learning rate $1e-3$ and the linear schedule decay rate 0.1 for every 3 epochs. The batch size for mini-batch optimization is 100 and we set an L2 regularization to $1e-5$ to avoid overfitting.

\begin{table}[t]
    \centering
    \caption{Performance compared with other baselines.}
    \vspace{-0.3cm}
    \scalebox{0.8}{
    \begin{tabular}{ccccccc}
         \toprule
         \multirow{2}*{Method}&\multicolumn{2}{c}{\textit{Yoochoose1/64}}&\multicolumn{2}{c}{\textit{Yoochoose1/4}}&\multicolumn{2}{c}{\textit{Diginetica}}\\
         &R@20&MRR@20&R@20&MRR@20&R@20&MRR@20\\
         \midrule
         POP&6.71&1.65&1.33&0.30&0.89&0.20\\
         S-POP&30.44&18.35&27.08&17.75&21.06&13.68\\
         Item-KNN&51.60&21.81&52.31&21.70&35.75&11.57\\
         BPR-MF&31.31&12.08&3.40&1.57&5.24&1.98\\
         FPMC&45.62&15.01&-&-&26.53&6.95\\
         GRU4REC&60.64&22.89&59.53&22.60&29.45&8.33\\
         NARM&68.32&28.63&69.73&29.23&49.70&16.17\\
         STAMP&68.74&29.67&70.44&30.00&45.64&14.32\\
         SR-GNN&70.57&30.94&71.36&31.89&50.73&17.59\\
         \midrule
         (basic session graph)\\
         FGNN-SG-Gated&70.85&31.05&71.50&32.17&51.03&17.86\\
         FGNN-SG-ATT&70.74&31.16&71.68&32.26&50.97&18.02\\
         FGNN-SG&$\bm{71.12}$&$\bm{31.68}$&$\bm{71.97}$&$\bm{32.54}$&$\bm{51.36}$&$\bm{18.47}$\\
         \midrule
         (BCS graph)\\
         FGNN-BCS-0&71.43&31.97&72.21&32.66&51.45&18.57\\
         FGNN-BCS-1&71.68&32.34&$\bm{72.53}$&$\bm{32.80}$&51.59&18.72\\
         FGNN-BCS-2&$\bm{71.75}$&$\bm{32.45}$&72.48&32.71&$\bm{51.67}$&18.69\\
         FGNN-BCS-3&71.52&32.31&72.33&32.68&51.63&$\bm{18.74}$\\
         \bottomrule
    \end{tabular}}
    \label{all-baseline}
\end{table}

\subsection{Comparison with Baseline Methods (RQ1)}
\label{rq1}
To demonstrate the overall performance of FGNN, we compare it with the baseline methods mentioned in Section~\ref{baselines} by evaluating their scores of R@20 and MRR@20. The overall results are presented in Table~\ref{all-baseline} with respect to all baseline methods and our proposed FGNN model. Due to the insufficient memory of hardware, we can not initialize FPMC on \textit{Yoochoose1/4} as~\cite{li2017neural}, which is not reported in Table~\ref{all-baseline}. For more detailed comparisons, in Table~\ref{tab:5and10}, we present the results of the most recent state-of-the-art methods for the dataset \textit{Yoochoose1/64} when $K=5$ and $10$.

\subsubsection{General Comparison by P@20 and MRR@20}
\label{sec:gen-com}
FGNN utilizes the multi layers of WGAT to easily convey the semantic and structural information between items within the session graph and applies the Readout function to decide the relative significance as the order of nodes in the graph to make the recommendation. According to the results reported in Table~\ref{all-baseline}, it is obvious that the proposed FGNN model outperforms all the baseline methods on all three datasets for both metrics, R@20 and MRR@20. In the experiment, we compare the basic session graph (FGNN-SG) and the BCS graph (FGNN-BCS-$n$). Especially, for FGNN-SG and FGNN-BCS-0, although they have the same structure of the session graph without extra nodes, FGNN-BCS-0 still gains better performance than FGNN-SG because the edge features also contain cross-session information for BCS graph. It is proved that our method achieves state-of-the-art performance on every dataset in both the methods of building session graphs. We also substitute the two key components, WGAT and the Readout function, with gated graph networks (FGNN-SG-Gated) and the self-attention (FGNN-SG-ATT) used by previous methods. Both of the variants gain improvements compared with previous models, which demonstrate the efficacy of the proposed WGAT and the Readout function respectively.

Compared with those traditional algorithms, e.g., POP and S-POP, which simply recommend items based on the frequencies of appearance, FGNN performs much better in overall. They tend to recommend fixed items, which leads to the failure of capturing the characteristics of different items and sessions. Taking BPR-MF and FPMC into consideration, which omit the session setting when recommending items, we can see that S-POP can defeat these methods as well because S-POP makes use of the session context information. Item-KNN achieves the best results among the traditional methods, although it only calculates the similarity between items without considering sequential information. At the even worse situation when the dataset is large, methods relying on the whole item set undoubtedly fail to scale well. All methods above achieve relatively poor results compared with the recent neural-network-based methods, which fully model the user's preference in the session sequence.

Different from the traditional methods mentioned above, all baselines using neural networks achieve a large performance margin. GRU4REC is the first to apply RNN-like units to encode the session sequence. It sets the baseline of neural-network-based methods. Although RNN is perfectly matched for sequence modeling, session-based recommendation problems are not merely a sequence modeling task because the user's preference is even changing within the session. RNN takes every input item equally importantly, which introduces bias to the model during training. For the subsequent methods, NARM and STAMP, both of which incorporate a self-attention over the last input item of a session, they both outperform GRU4REC in a large margin. They both use the last input item as the representation of short-term user interest. It proves that assigning different attention to different inputs is a more accurate modeling method for session encoding. Looking into the comparison between NARM, combining RNN and attention mechanism, and STAMP, a complete attention setting, there is a conspicuous gap of performance that STAMP outperforms NARM. This further demonstrates that directly using RNN to encode the session sequence can inevitably introduce bias to the model, which the attention can completely avoid.

SR-GNN uses a session graph to represent the session sequence, followed by a gated graph layer to encode items. In the final stage, it again uses a self-attention the same as STAMP to output a session embedding. It achieves the best result compared to all the methods mentioned above. The graph structure is shown to be more suitable than the sequence structure, the RNN modeling, or a set structure, the attention modeling.

\begin{table}[t]
    \centering
    \caption{Performance when $K=5$ and $10$ for \textit{Yoochoose1/64}.}
    \vspace{-0.3cm}
    \scalebox{0.8}{
    \begin{tabular}{ccccc}
         \toprule
         \multirow{2}*{Method}&\multicolumn{4}{c}{\textit{Yoochoose1/64}}\\
         &R@5&MRR@5&R@10&MRR@10\\
         \midrule
         NARM&44.34&26.21&57.50&27.97\\
         STAMP&45.69&27.26&58.07&28.92\\
         SR-GNN&47.42&28.41&60.21&30.13\\
         FGNN-SG&$\bm{48.23}$&$\bm{29.16}$&$\bm{60.97}$&$\bm{30.85}$\\
         \midrule
         FGNN-BCS-0&48.30&29.41&61.07&30.94\\
         FGNN-BCS-1&$\bm{48.41}$&29.48&61.14&$\bm{31.12}$\\
         FGNN-BCS-2&48.36&$\bm{29.50}$&$\bm{61.19}$&31.09\\
         FGNN-BCS-3&48.32&29.45&61.15&31.06\\
         \bottomrule
    \end{tabular}}
    \label{tab:5and10}
\end{table}

\subsubsection{Higher Standard Recommendation with $K=5,10$}
\label{sec:high-std}
For more detailed results in Table~\ref{tab:5and10}, FGNN also achieves the best results with a higher standard of the top-5 and top-10 recommendations. The proposed FGNN model outperforms all baseline methods above. It has a more accurate node-level encoding tool, WGAT, to learn more representative features and a Readout function, to learn an inherent order of nodes in the graph to avoid the entire random order of items. According to the result, it is demonstrated that a more accurate session embedding is obtained by FGNN to make effective recommendations, which proves the efficacy of the proposed FGNN.

\subsection{Comparison with Different Session Graph Generation (RQ2)}
\label{rq2}
Different methods of generating a graph for model forward computation can result in different levels of cross-session information incorporation. Basically, we can build a session graph for each session graph without any cross-session information. We refer to this setting as FGNN-SG. For the BCS graph setting, we can obtain $0$-hop neighbor situation $G_{\text{BCS}-0}$ with the same structure as the basic session graph but different edge weights. We refer to this setting as FGNN-BCS-0. Similarly, we also have comparisons between different $n$-hop neighbors. Therefore, we conduct experiments with FGNN-BCS-1, FGNN-BCS-2 and FGNN-BCS-3 as well.

\subsubsection{General Comparison}
\label{sec:gen-com-bcs}
In Table~\ref{all-baseline} and~\ref{tab:5and10}, it is clear that all BCS graph based methods outperform the FGNN-SG method introduced by the previous version of this paper in the aspect of both R@20 and MRR@20 for all datasets. Basic session graph setting shows worse results compared to all BCS graph methods, which means apparently that the cross-session information can be represented by the BCS graph and learned by the following GNN layers. Looking into the difference for the choice of the neighborhood, when it reaches the 3-hop, the performance does not increase as the nodes become more and more. This peak of the performance is because of the GNN layer can not go as deep as the convolutional layer. Meanwhile, the more nodes introduce more information and noise to learn by the model, which is the case being too difficult for the shallow GNN model.

\subsubsection{Performance on Session with Different Lengths}
In addition to the overall performance of the BCS graph module, we further conduct experiments to analyze the detailed performance for sessions with different lengths. Following the previous work~\cite{wu2018session,Liu18STAMP}, sessions in \textit{Yoochoose 1/64} are separated into two groups, i.e., short sessions and long sessions. Short sessions indicate that the length of sessions is less than or equal to 5, while sessions longer than 5 are categorized as long sessions. Length 5 is the closest to the average length of total sessions. $70.1\%$ of \textit{Yoochoose1/64} are short sessions and $29.9\%$ are long sessions.

\begin{figure}[t]
    \centering
    \includegraphics[width=0.5\linewidth]{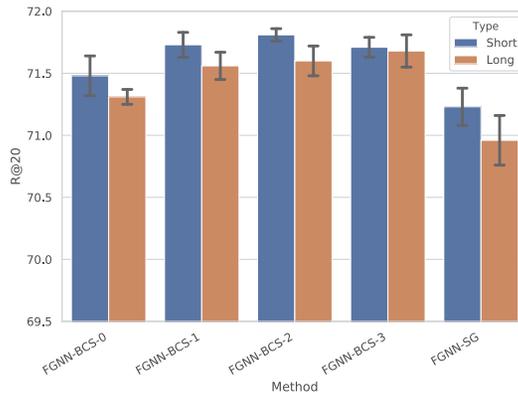}
    \vspace{-0.4cm}
    \caption{R@20 for Short and Long sessions with different methods to build the session graph. The blue bars stand for short sessions and the orange ones for long sessions.}
    \vspace{-0.4cm}
    \label{fig:p20-long-short_graph}
\end{figure}

\begin{figure}[t]
    \centering
    \includegraphics[width=0.5\linewidth]{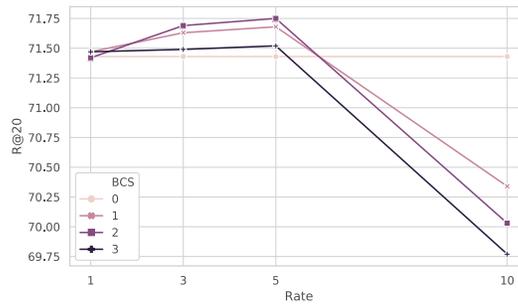}
    \vspace{-0.4cm}
    \caption{R@20 index for different sampling rate ranging in $\{1,3,5,10\}$ and neighboring hop for BCS graph.}
    \label{fig:p20-bcs-line}
\end{figure}

In Fig.\ \ref{fig:p20-long-short_graph}, the results of the performance on different sessions with the basic graph and the BCS graph are presented on the R@20 metric on \textit{Yoochoose 1/64} dataset. For the short session, FGNN-BCS-2 performs the best while for the long session, FGNN-BCS-3 shows its superiority on other methods. Compared with all BCS graph methods, FGNN-SG performs worse on both the long and short sessions. Therefore, we can draw a conclusion that incorporating the cross-session information enables the model to gain improvements from sessions of different lengths. For short sessions, the BCS graph with 2-hop neighbors achieves the best result due to the better representation learning ability for shallow GNN models. In contrast, the BCS graph with 3-hop neighbors performs better as the sessions grow longer. It may be because the Readout function can make use of a longer original session when more nodes are from other sessions. Since there are much shorter sessions in the datasets, the overall performance is dominated by the short sessions as shown in Table~\ref{all-baseline} and~\ref{tab:5and10}.

\subsubsection{Performance with Different Sample Rates of BCS Graph}
AS the choice of neighboring nodes goes more and more beyond from 0-hop to 3-hop, the session graph becomes larger and larger while the performance shown in the previous tables and graphs begins to decrease. According to the analysis above, the wider of the neighborhood is chosen, the noisier the session graph will be. And the depth of the GNN model does not grow as convolutional neural networks do. Our model is restricted to gain further better results due to the reasons above. As a result, in the larger BCS graph, we conduct random sampling on the neighboring nodes when building the BCS graph. For the BCS-2 and BCS-3 cases, we perform random sampling ranging in $\{1,3,5,10\}$ under the distribution based on the edge weight connected to the current node. For the BCS-0 situation, because there is no neighbor node included, the sampling procedure does not work here.

In Fig.\ \ref{fig:p20-bcs-line}, the results with different sampling rates of different neighboring choices are shown. When incorporating different neighboring nodes, the performance is best at the sampling rate at 5 for different choices of the BCS graph. When the sampling rate is 10, the performance decreases for every BCS graph. A large number of neighboring nodes can bring noisy data to the session representation learning, which is harmful to the model. An adequate choice of the neighboring nodes can boost the performance with the cross-session information.

\subsection{Comparison with Other GNN Layers (RQ3)}
\label{rq3}
To efficiently convey information between items in a session graph, we propose to use WGAT, which suits the situation of the session better. As mentioned above, there are many different GNN layers that can be used to generate node embeddings, e.g., GCN~\cite{kipf2017semi}, GAT~\cite{velickovic2018graph} and gated graph networks~\cite{li2015gated,wu2018session}. To prove the usefulness of WGAT, we substitute all three WGAT layers with GCN, GAT and gated graph networks respectively in our model. For GCN and GAT, they both initially work for the unweighted and undirected graph, which is not the same setting as the proposed session graph. To make both of them work on the session graph, we directly convert the session graph into undirected by replacing the originally directed edges with undirected ones, i.e., reverse the source and target nodes of edges. And we simply omit the original weight of edges and set all connections between nodes with the same weight 1. For the other one, Gated graph networks, it can work with the session graph setting in its original form without any modification on the session graph.

\begin{figure}[t]
    \centering
    \subfigure[R@20 index.]{
    \label{fig:p20-wgat}
    \includegraphics[width=0.47\linewidth]{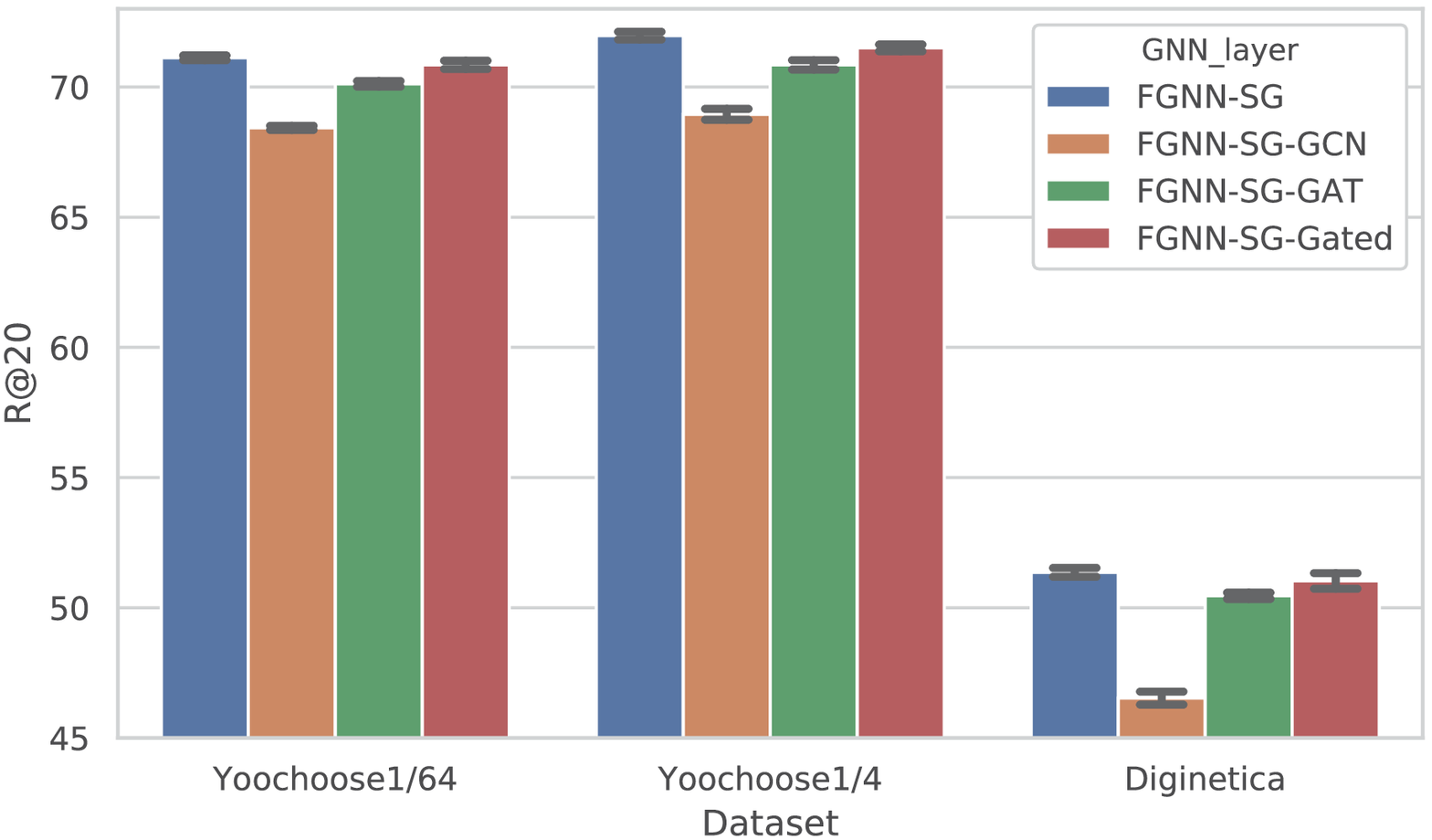}
    }
    \subfigure[MRR@20 index.]{
    \label{fig:mrr20-wgat}
    \includegraphics[width=0.47\linewidth]{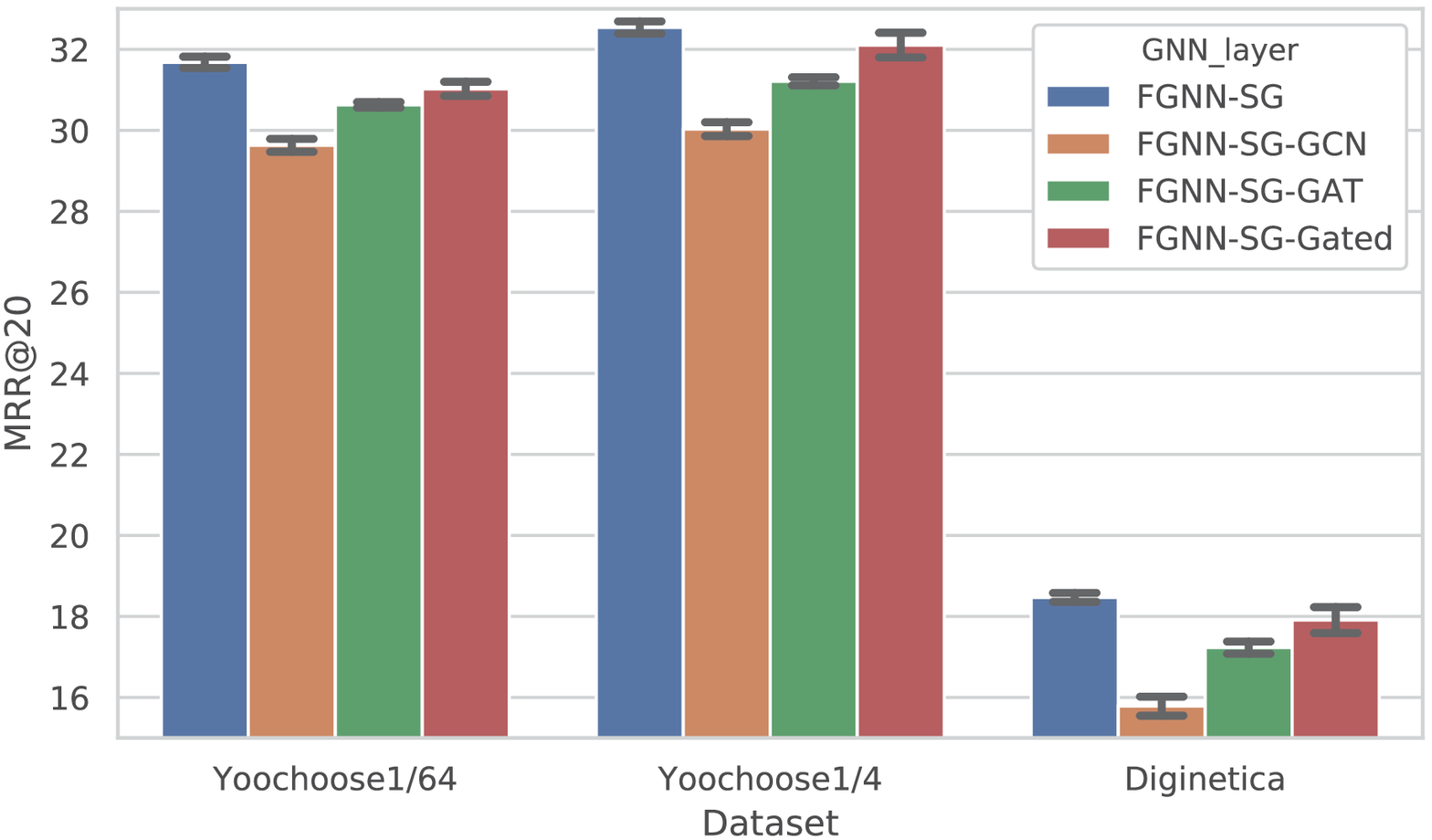}
    }
    \vspace{-0.4cm}
    \caption{Results with different GNN layers.}
    \vspace{-0.5cm}
\end{figure}

\subsubsection{General Comparison}
In Fig.\ \ref{fig:p20-wgat} and Fig.\ \ref{fig:mrr20-wgat}, results of different GNN layers are shown with R@20 and MRR@20 indices. FGNN is the model proposed in this work, which achieves the best performance. WGAT is more powerful than other GNN layers in the session-based recommendation. GCN and GAT are not able to capture the direction and the explicit weight of edges, resulting in performing worse than WGAT and gated graph networks, which hold the ability to capture this information. Between WGAT and gated graph networks, WGAT performs better because of the stronger ability of representation learning.

\begin{figure}[t]
    \centering
    \includegraphics[width=0.5\linewidth]{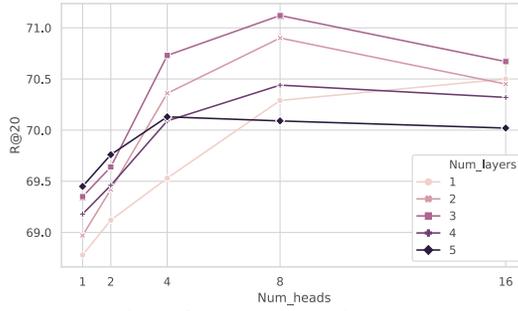}
    \vspace{-0.4cm}
    \caption{R@20 index for different number of layers and heads for WGAT.}
    \vspace{-0.5cm}
    \label{fig:p20-wgat-line}
\end{figure}

\begin{figure*}[t]
    \centering
    \subfigure[R@20 index.]{
    \label{fig:p20-readout}
    \includegraphics[width=0.47\linewidth]{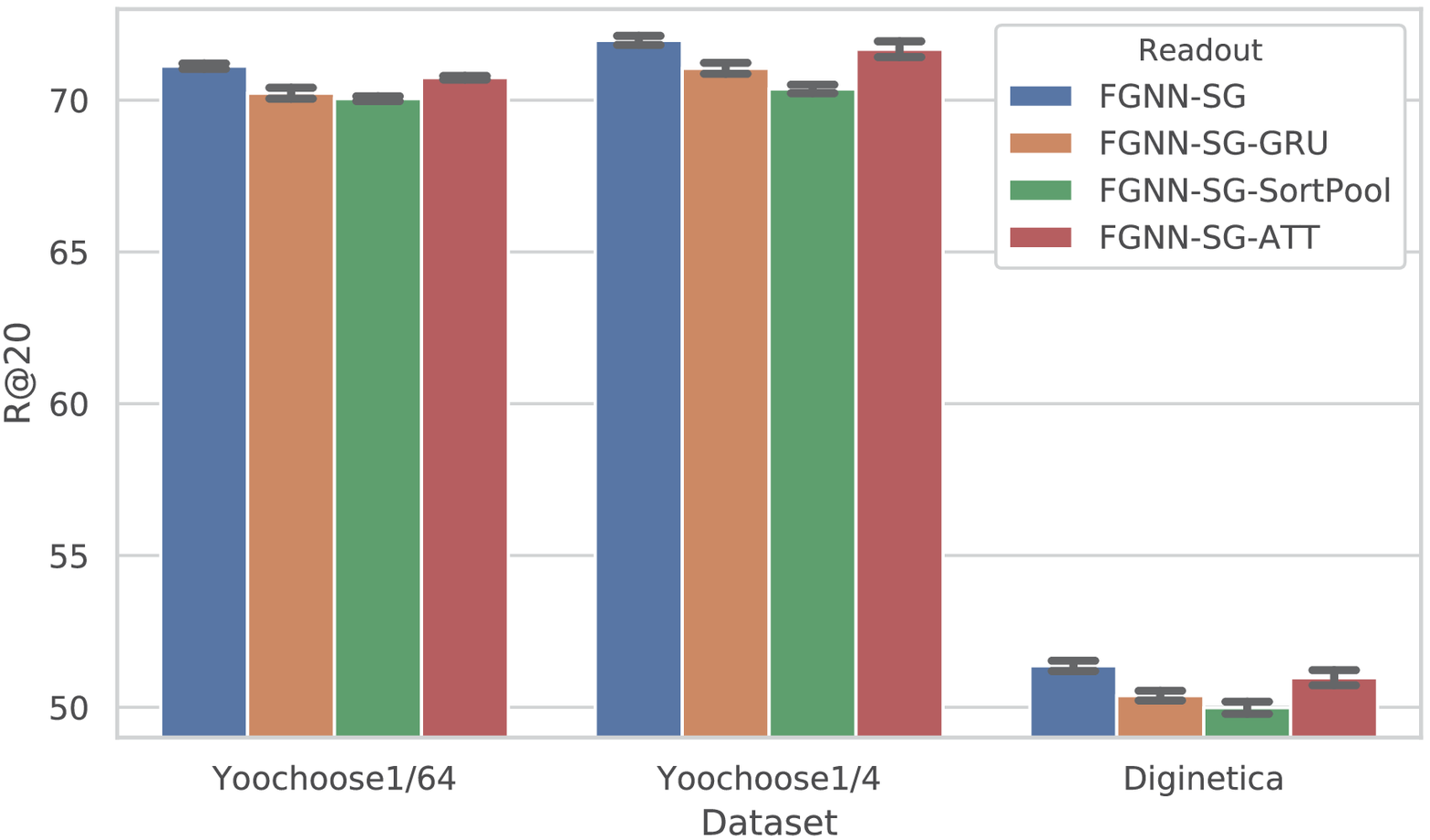}
    }
    \subfigure[MRR@20 index.]{
    \label{fig:mrr20-readout}
    \includegraphics[width=0.47\linewidth]{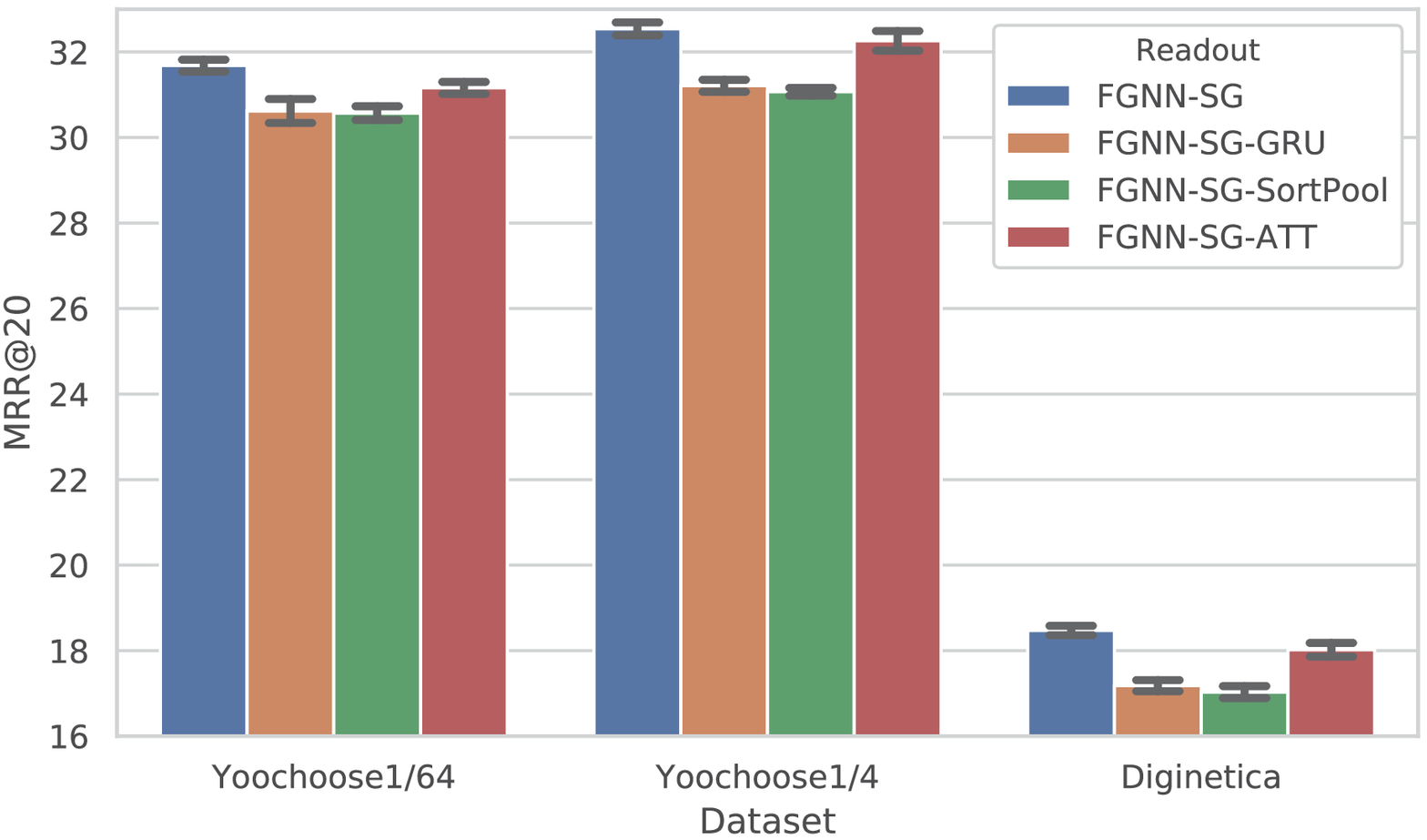}
    }
    \vspace{-0.4cm}
    \caption{Results with different aggregation functions.}
\end{figure*}

\subsubsection{Performance of Different Numbers of Head and Layer}
For the study of WGAT, we test how the number of layers and heads affect the R@20 index performance on \textit{Yooshoose1/64}. In Fig.\ \ref{fig:p20-wgat-line}, we report the experiment results of different number of layers ranging in $\{1,2,3,4,5\}$ and heads ranging in $\{1,2,4,8,16\}$. It shows that stacking three WGAT layers with eight heads performs the best. Lower results are shown for smaller models for the reason that the capacity of them is too low to represent the complexity of the item dependency relationship. According to the tendency of results of larger models, it is too complicated to train the model and the overfitting does harm to the final performance.

\subsection{Comparison with Other Graph Embedding Methods (RQ4)}
\label{rq4}
Different approaches for generating the session embedding after obtaining the node embeddings stand for different emphasis of the input items. The Readout function proposed in this work learns an inherent order of the nodes by the query vector, which indicates the relatively different impact on the user's preference along with the item dependency. To prove the superiority of our Readout function, we replace the Readout function with other session embedding generators:
\begin{itemize}
    \item \textbf{FGNN-SG-ATT} We apply the widely-used self-attention of the last input item. It directly considers the last input item as the short-term reference and all other items as the long-term reference.
    \item \textbf{FGNN-SG-GRU} To compare the inherent order learned by our Readout function, we use GRU to directly make use of the input session sequence order.
    \item \textbf{FGNN-SG-SortPool} SortPooling is introduced by Zhang et al.~\cite{zhang2018end} to perform a pooling on graph level by sorting the features of nodes. This sorting can be viewed as a kind of order as well.
\end{itemize}

\subsubsection{General Comparison}
In Fig.\ \ref{fig:p20-readout} and Fig.\ \ref{fig:mrr20-readout}, results of different methods for graph level embedding generation are presented for all three datasets with the R@20 and MRR@20 indices. It is obvious that the proposed Readout function achieves the best result. For FGNN-GRU and FGNN-SortPool, they both contain an order but it is too simple to capture the item dependency relationship. FGNN-GRU uses GRU to encode the session sequence with the input order. Such a setting is similar to common RNN-based methods. As a consequence, it performs worse than the attention-based method FGNN-ATT-OUT, which takes both the short-term and the long-term preference into consideration. As for FGNN-SortPool, it sorts the nodes based on WL colors from previous multiple layers of computations. Although it does not simply rely on the input order of the session sequence, the order for the nodes is set according to the relative scale of the features. For the best performance, our Readout function learns the order of the item dependency relationship, which is different from using the time order or the hand-crafted split of long-term and short-term preference. The results prove that there is a more accurate order for the model to make a more accurate recommendation.

\subsection{Mask-Readout Compared with Readout}
\label{rq5}
In this section, we conduct experiments using the BCS graph setting and compare the experimental results with Mask-Readout or Readout. When the session is converted into a basic session graph or a BCS-$0$ graph, there will be no difference between Mask-Readout and Readout because the choices of nodes are the same for both of them. Overall, the experiments are conducted on BCS-$1$,-$2$ and -$3$ graphs. We use \textbf{FGNN-BCS-$n$-Readout} ($n$ indicates different neighbors) to denote the methods which use the previous Readout function and \textbf{FGNN-BCS-$n$} to denote the methods which use Mask-Readout.

\begin{figure}[t]
    \centering
    \subfigure[R@20 index.]{
    \label{fig:p20-cent}
    \includegraphics[width=0.47\linewidth]{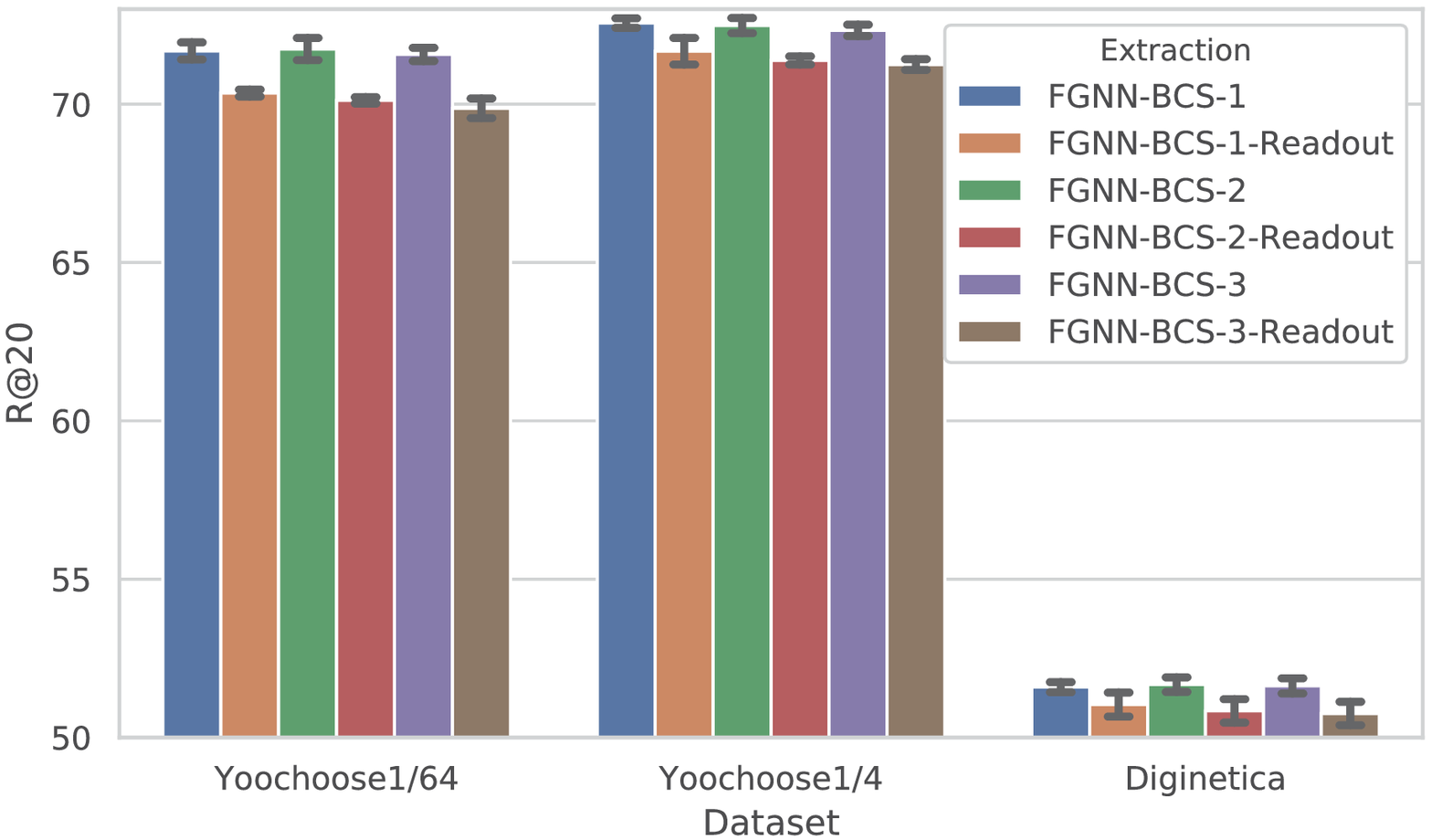}
    }
    \subfigure[MRR@20 index.]{
    \label{fig:mrr20-cent}
    \includegraphics[width=0.47\linewidth]{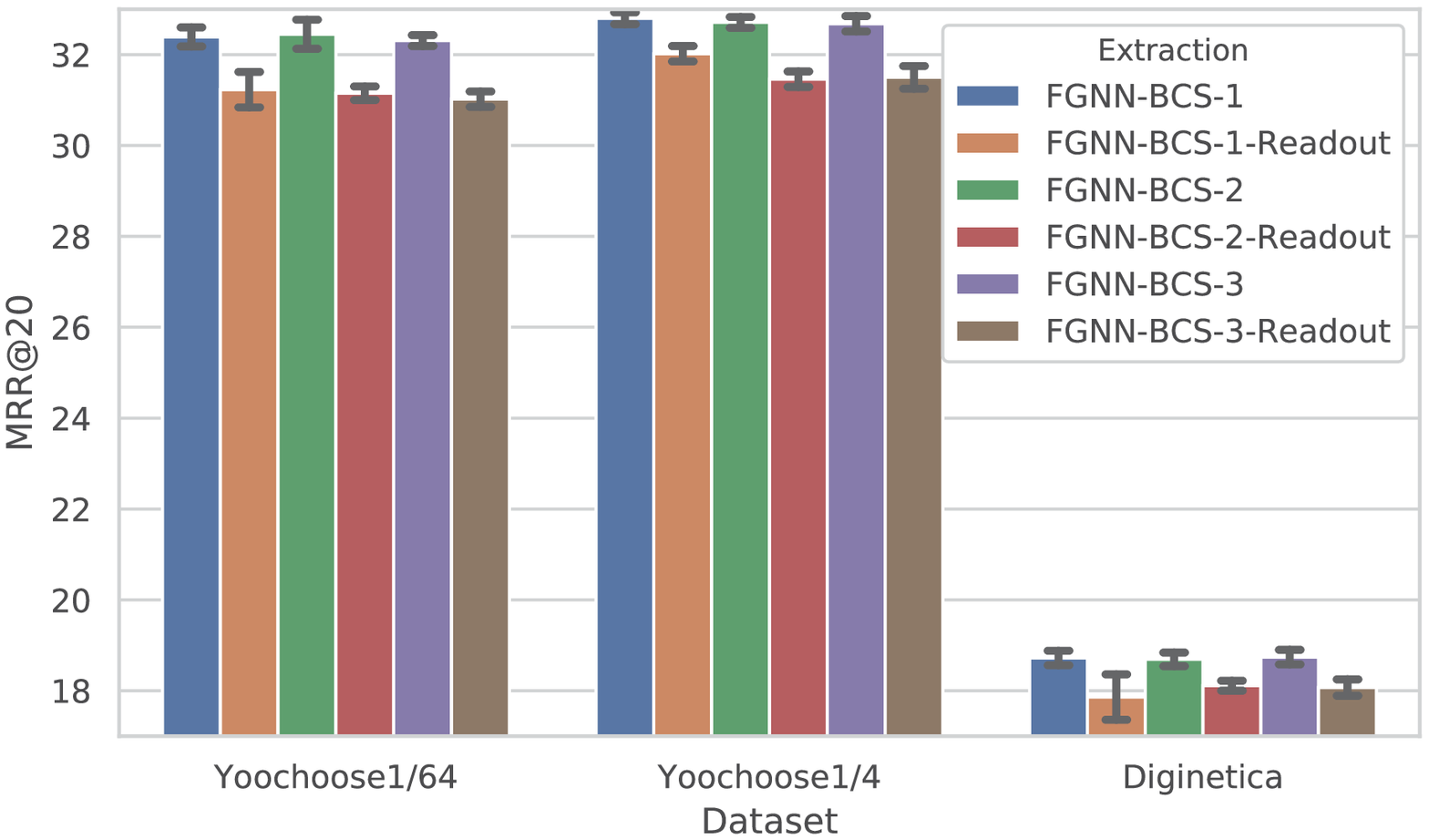}
    }
    \vspace{-0.4cm}
    \caption{Results with FGNN using Mask-Readout or Readout. }
\end{figure}

Fig.\ \ref{fig:p20-cent} and~\ref{fig:mrr20-cent} demonstrate the result of how different graph representation extraction methods perform on all three datasets with the metrics of R@20 and MRR@20. According to the results, it is clear that Mask-Readout outperforms the previous Readout in all situations, which means that Mask-Readout is more suitable for the BCS graph. Compared with Readout, Mask-Readout only cares about the nodes that are in the individual session rather than other sessions. Consequently, Cent-Read prevents the model from losing the session information, which becomes back to the simple dependency information when processing with Readout. Overall, Mask-Readout helps the model to balance the cross-session information and the individual session information.

Take a deeper look at Fig.\ \ref{fig:p20-cent} and~\ref{fig:mrr20-cent}, when the Readout is applied, as the neighbors in the BCS graph grow, the performance generally becomes worse. This situation indicates that the more cross-session information is incorporated, the more easily our model will be distracted from the individual session. But when Mask-Readout is used instead of Readout, this phenomenon weakens. 

\section{Conclusion}
\label{conclusion}
This paper studied the problem of the session-based recommendation on anonymous sessions in the aspect of the complicated item dependency and the cross-session information. We found that a sequence or a random set of items are insufficient to capture the relation between items. To learn the complicated item dependency, we first represented each session as a graph and then proposed an FGNN model to perform graph convolution on the session graph. Furthermore, because of the data sparsity issue of the anonymous session, it is helpful to make use of the cross-session information. To exploit and incorporate the cross-session information, we further designed a BCS graph to connect different sessions and used a Mask-Readout function to generate a more expressive session embedding with the cross-session information. Empirically, the experimental results from two large-scale benchmark datasets validated the superiority of our solution compared with state-of-the-art techniques.

\bibliographystyle{ACM-Reference-Format}
\bibliography{TOIS-2019-0129.bib}

\end{document}